# Solution enthalpies of barium cerates co-doped by rare-earth elements and indium

N.I. Matskevich[1], Th. Wolf[2], A.N. Semerikova[1], O.I. Anyfrieva[1], M.Yu. Matskevich[1], I.V. Vyazovkin[1]

[1]Nikolaev Institute of Inorganic Chemistry SB RAS, Novosibirsk 630090, Russia

[2]Karlsruhe Institute of Technology, Institute of Solid State Physics, Karlsruhe, Germany

## Abstract

The dissolution enthalpies of $BaCe_{0.7}Ho_{0.2}In_{0.1}O_{2.85}$ and $BaCe_{0.7}Sm_{0.2}In_{0.1}O_{2.85}$ were measured by method of solution calorimetry in 1M hydrochloric acid with adding 0.1 M KI. The obtained data were compared with earlier measured data on dissolution enthalpies of $BaCe_{0.7}Nd_{0.2}In_{0.1}O_{2.85}$, $BaCe_{0.7}Gd_{0.2}In_{0.1}O_{2.85}$. The dependence of dissolution enthalpies from lanthanide radius was constructed. As was shown the relationship is not linear.

## 1. Introduction

Compounds on the basis of barium cerate doped by rare-earth and other elements are perspective materials for using in fuel cells, sensors, batteries, electro catalysis devices, pigments, etc. The general formula is $BaCe_{1-x}M_xO_y$ (M – metals of third or other group).

When the compounds are used as electrolytes of fuel cell, it is important to know the homogeneity field and compounds stability, in particularly, thermodynamic stability. Constructed dependences "thermodynamic properties – structure" allow us to understand the tendency of properties changing and then to optimize the synthesis conditions of employed compounds.

For successful application of compounds on the basis of barium cerates doped by rare-earth elements as fuel cell electrolytes, it is necessary to extend the homogeneity field. The limit of homogeneity field for compounds $BaCe_{1-x}R_xO_y$ (R - rare earth element) is typically 20%. It was found in papers [1-3] that doping indium instead of rare earth elements the homogeneity region can be expanded in 4 times, namely up to to 80%. I.e. it is possible to obtain the compound with general formula $BaCe_{1-x}In_xO_y$ (x = 0-0.8). However as it was found the conductivity of indium doped cerates is less than cerates with rare earth elements.

To the one hand to increase the homogeneity region and thermodynamic stability, and on the other hand do not reduce the conductivity the strategy of "co-doping" by indium and rare-earth elements has been suggested in papers [4-10]. As it was shown in paper [4] the best composition having good conductivity and thermodynamic stability is following: $BaCe_{0.7}Y_{0.2}In_{0.1}O_{2.85}$. It is advisable to study the compounds with the same composition with other rare earth elements.

In this paper we synthesized compounds $BaCe_{0.7}Ho_{0.2}In_{0.1}O_{2.85}$, $BaCe_{0.7}Sm_{0.2}In_{0.1}O_{2.85}$ and measured their dissolution enthalpy in hydrochloric acid. In addition we performed the comparison of measured dissolution enthalpies with dissolution enthalpies for compounds $BaCe_{0.7}Nd_{0.2}In_{0.1}O_{2.85}$ and $BaCe_{0.7}Gd_{0.2}In_{0.1}O_{2.85}$ which were measured earlier by us. We also constructed the dependence of dissolution enthalpies from rare earth element radius.

## 2. Experimental part

The synthesis of $BaCe_{0.7}Ho_{0.2}In_{0.1}O_{2.85}$, $BaCe_{0.7}Sm_{0.2}In_{0.1}O_{2.85}$ was performed according to the reaction: $BaCO_3 + 0.5x\ (M,R)_2O_3 + (1-x)CeO_2 = BaCe_{1-x}(M,R)_xO_{3-x/2} + CO_2$ (M = R, In). The following reagents were using for preparation: $BaCO_3$ (>99%, MERCK), $CeO_2$ (99.99%, Johnson Matthey GmbH, Alfa Products), $In_2O_3$ (82.1% In, Johnson Matthey, Materials Technology UK), $Ho_2O_3$ (99.99%, Vetron), $Sm_2O_3$ (99.99%, Johnson Matthey Company).

The technology of synthesis was the following. Starting reagents were treated before synthesis at 600 ºC ($In_2O_3$), 800 ºC ($R_2O_3$) up to constant weight. Then they were mixed in an agate mortar, ground in a planetary ball mill, pressed and heat treated in air according to the following regime: 800 ºC for 16 h, 1100 ºC for 10-16 h, 1440-1450 ºC for 40 h, 800 ºC for 4-10 h. X-ray analysis showed that all samples were single phase. X-ray pattern for $BaCe_{0.7}Ho_{0.2}In_{0.1}O_{2.85}$ is presented in Fig. 1. Structure is orthorhombic, space group *Pmcn.*

We have increased the homogeneity limit in following new compounds $BaCe_{0.7}Ho_{0.2}In_{0.1}O_{2.85}$, $BaCe_{0.7}Sm_{0.2}In_{0.1}O_{2.85}$ by the co-doping method up to x = 0.3.

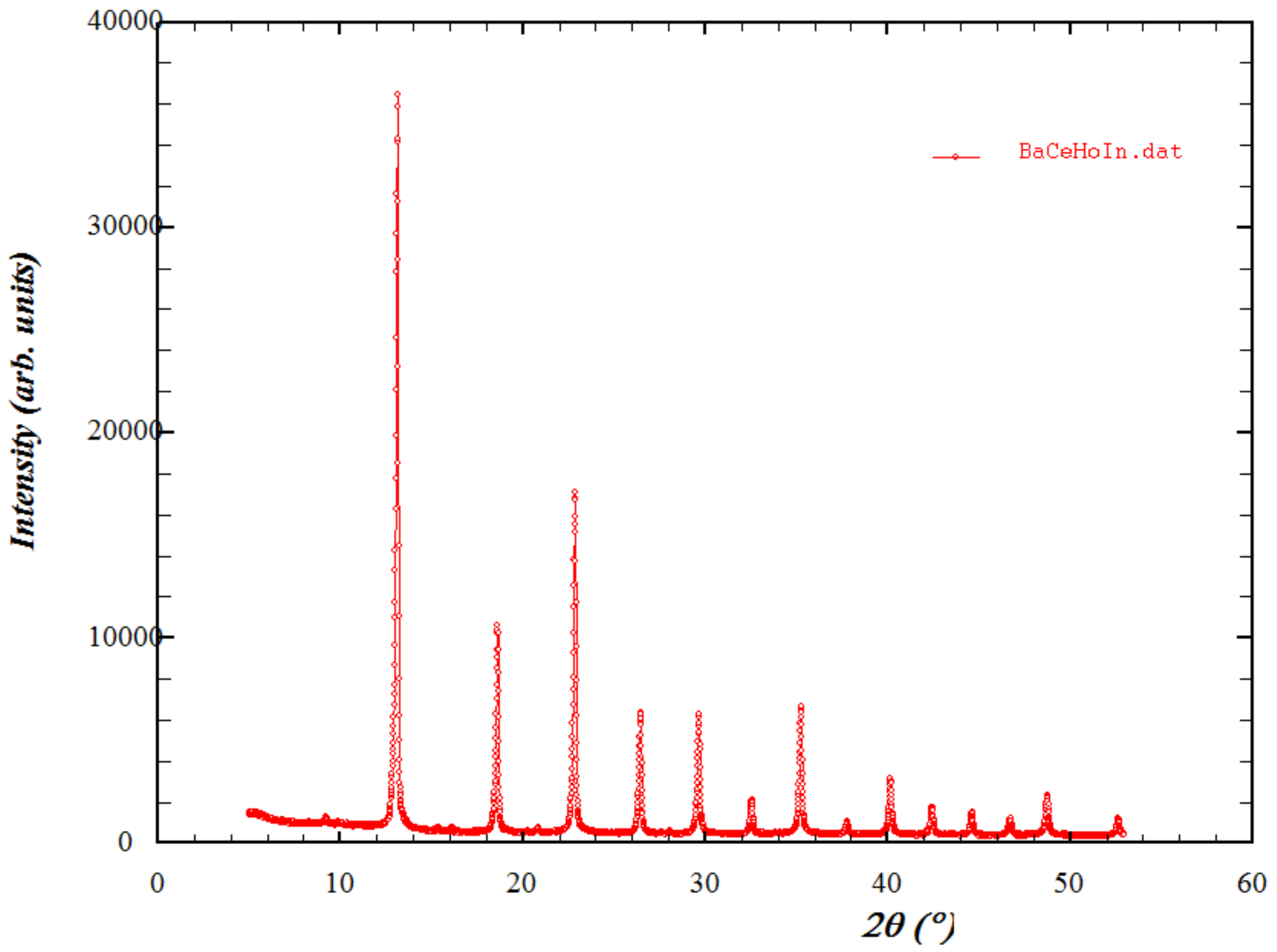

Figure 1. X-ray pattern for $BaCe_{0.7}Ho_{0.2}In_{0.1}O_{2.85}$

Solution calorimetry with 1 M HCl + 0.1 M KI as solvent was chosen as a method to determine the solution enthalpies of $BaCe_{0.7}Ho_{0.2}In_{0.1}O_{2.85}$, $BaCe_{0.7}Sm_{0.2}In_{0.1}O_{2.85}$. Experiments were performed in an automatic solution calorimeter with isothermal shield. Construction of calorimeter and procedure of carrying out experiments were described earlier [11-13]. The main part of the calorimeter was a glass Dewar vessel (250 ml). The thermometer, calibration heater, mixer and the device to break the ampoules were mounted on the lid closing the Dewar vessel. A device was created to connect the calorimeter heater and thermometer with a computer. The reproducibility of the heat equivalent of the calorimeter was 0.03%. Dissolution of a standard substance (KCl) was performed to check the precision of the calorimeter. The obtained dissolution heat of KCl (17.529 ± 0.009 kJ/mol) is in a good agreement with the value recommended in the literature.  The experiments were performed at 298.15 K. The amounts of substances used were about 0.04 g.

## 3. Results and discussion

We measured the dissolution enthalpies for $BaCe_{0.7}Ho_{0.2}In_{0.1}O_{2.85}$, $BaCe_{0.7}Sm_{0.2}In_{0.1}O_{2.85}$ as following values: $\Delta_{sol}H^o$ ($BaCe_{0.7}Ho_{0.2}In_{0.1}O_{2.85}$, 298.15 K) = –371.91 ± 2.67 kJ/mol; $\Delta_{sol}H^o$ ($BaCe_{0.7}Sm_{0.2}In_{0.1}O_{2.85}$, 298.15 K) = –353.51 ± 4.36 kJ/mol. The dissolution enthalpies of $BaCe_{0.7}Ho_{0.2}In_{0.1}O_{2.85}$, $BaCe_{0.7}Sm_{0.2}In_{0.1}O_{2.85}$ were calculated as average values of six experiments. Errors were calculated for the 95% confidence interval using the Student coefficient.

Earlier [7-8] we measured solution enthalpies for $BaCe_{0.7}Nd_{0.2}In_{0.1}O_{2.85}$, $BaCe_{0.7}Gd_{0.2}In_{0.1}O_{2.85}$ compounds as following values: $\Delta_{sol}H^o$ ($BaCe_{0.7}Nd_{0.2}In_{0.1}O_{2.85}$, 298.15 K) = –354.17 ± 3.21 kJ/mol; $\Delta_{sol}H^o$ ($BaCe_{0.7}Gd_{0.2}In_{0.1}O_{2.85}$, 298.15 K) = –331.58 ± 8.75 kJ/mol.

Below we constructed the dependence of solution enthalpies for investigated compounds from radius of rare-earth elements (Fig. 2). The radius was taken from Ref. [14]: $Nd^{+3}$ = 0.0983 nm; $Sm^{+3}$ = 0.0958 nm; $Gd^{+3}$ = 0.0938 nm; $Ho^{+3}$ = 0.0901 nm.

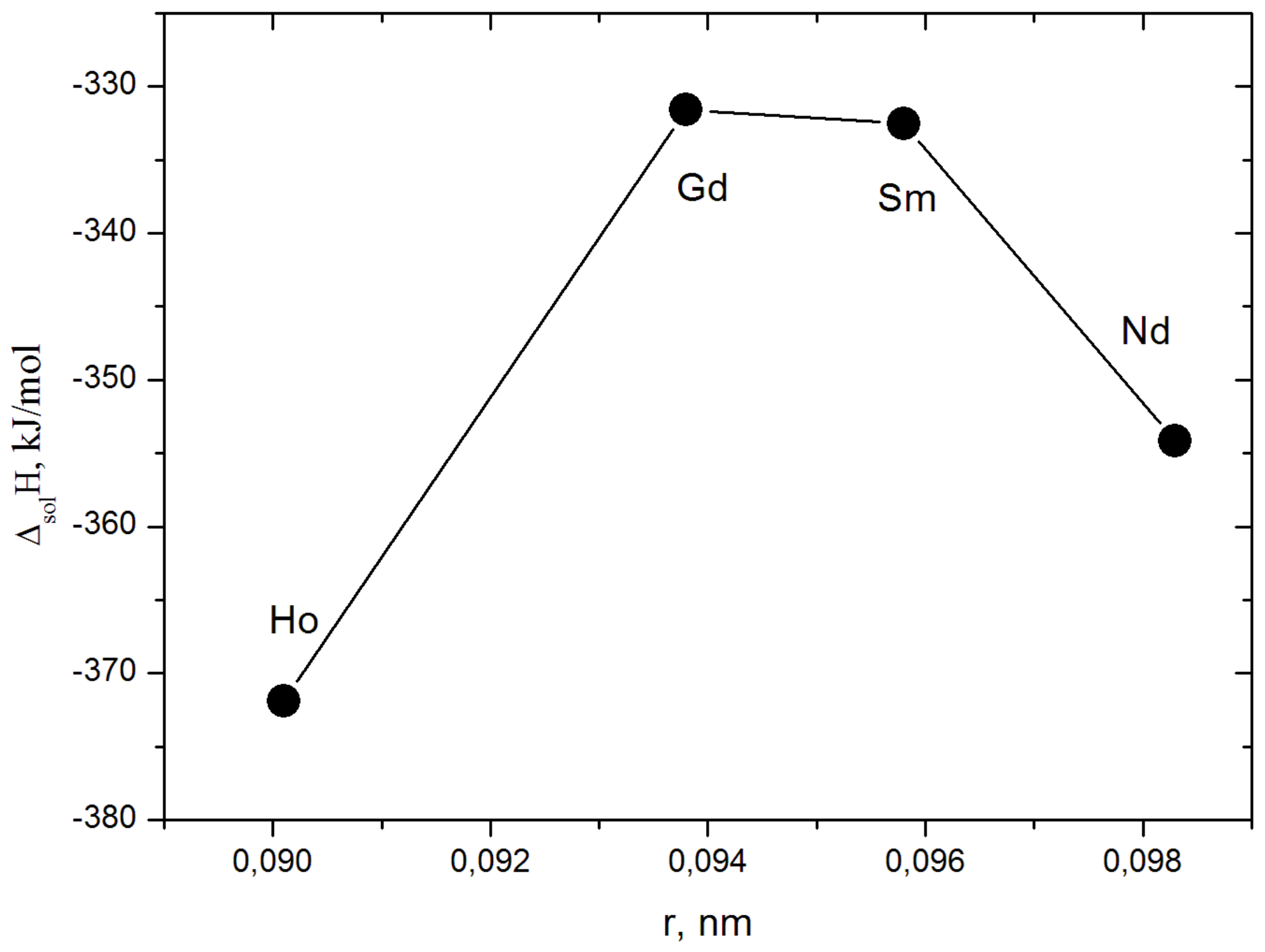

Figure 2. Dependence of $\Delta_{sol}H^o$ for $BaCe_{0.7}R_{0.2}In_{0.1}O_{2.85}$ from radius of rare-earth element

As it is possible to see, the dependence is not linear.

Earlier [15] the solution enthalpies of binary oxides ($Nd_2O_3$, $Sm_2O_3$, $Gd_2O_3$ and $Ho_2O_3$) were measured. The dependence of solution enthalpies for binary oxides from radius of rare-earth elements is presented in Fig. 3.

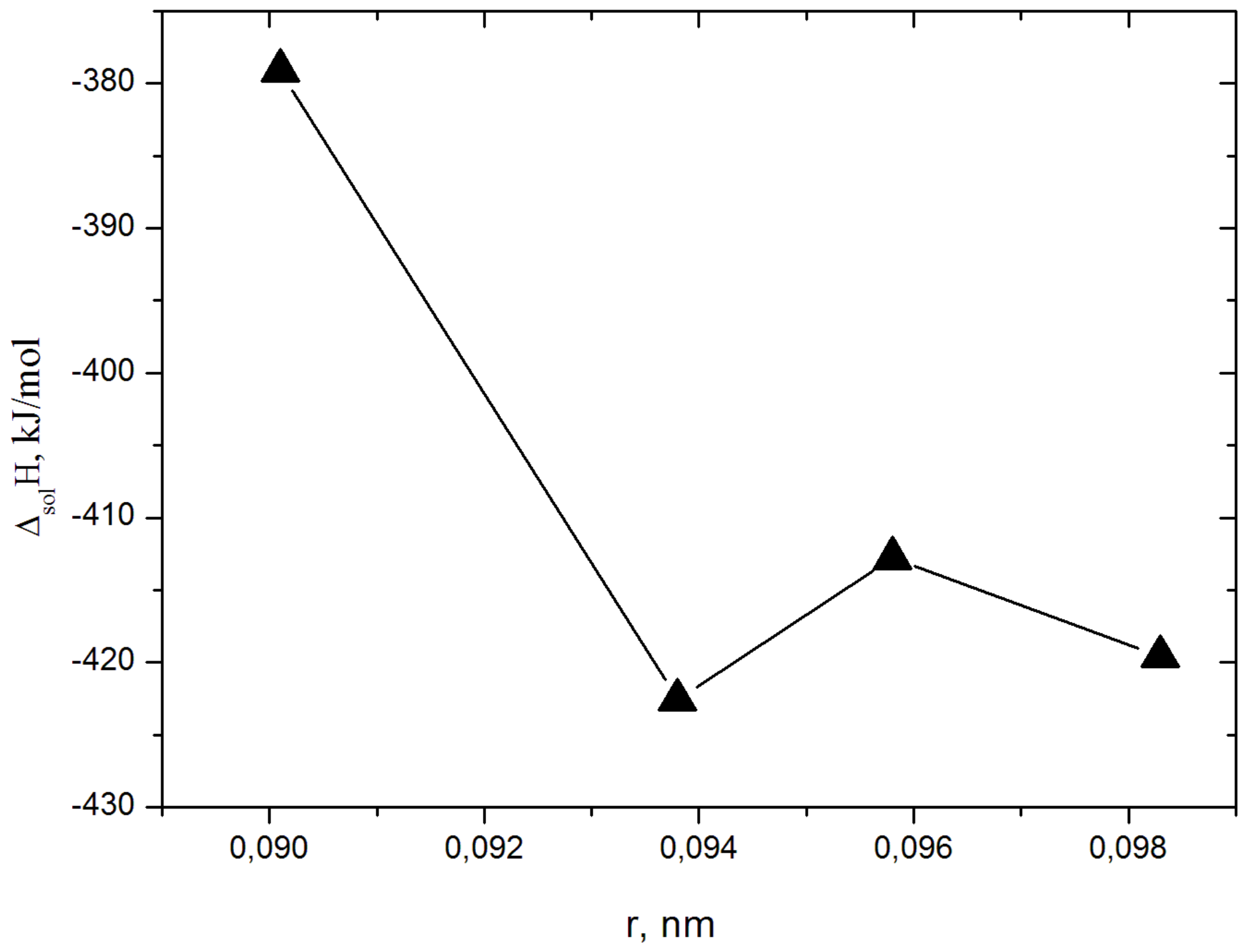

Figure 3. Dependence of $\Delta_{sol}H^o$ for binary oxides from radius of rare-earth element

As it is possible to see the dependence is not linear as well.

**Acknowledges**

This work is supported by Karlsruhe Institute of Technology, RFBR (project 16-08-00226_a) and Government Task for Nikolaev Institute of Inorganic Chemistry SB RAS.